\documentclass[aps,amsfonts,amssymb,amsmath, draft, nofootinbib]{revtex4}
\usepackage{bm,exscale, amsmath, mathrsfs, dsfont,accents}
\usepackage{times}
\usepackage{epsfig}
\usepackage[T1]{fontenc}
\usepackage{colordvi}

\newcommand{\La}{\ensuremath{\mathcal L}}

\begin{document} 
\title{Gauge invariant coupling of fields to torsion: a string inspired model}\author{Srijit Bhattacharjee}\email{srijit.bhattacharjee@saha.ac.in}\affiliation{Theory Group, Saha Institute of
Nuclear Physics, Kolkata 700064,India} 
\author{Ayan Chatterjee}\email{achatterjee@imsc.res.in}
\affiliation{The Institute of Mathematical Sciences, CIT Campus, Taramani, Chennai-600113, India}
 
\begin{abstract} 
In a consistent heterotic string theory, the Kalb-Ramond field, which is the source of spacetime torsion,
is augmented by Yang-Mills and gravitational Chern-Simons terms.
When compactified to $4$-dimensions and in the field theory limit, such additional
terms give rise to interactions with interesting astrophysical predictions like rotation of plane 
of polarization for electromagnetic and gravitational waves. On the other hand,
if one is also interested in coupling $2$ or $3$-form (Abelian or non-Abelian) gauge fields to torsion,
one needs another class of interaction. In this paper,
we shall study this interaction and offer some astrophysical and cosmological predictions. We also comment on the 
possibility of such terms in loop quantum gravity
where, if the Barbero-Immirzi parameter is promoted to a field, acts as a source for torsion. 
\end{abstract}
\maketitle
\section{Introduction}

The low energy physics of particle interactions is satisfactorily 
described by the standard model and general relativity. At higher
energies available at the early universe or at astrophysical processes,
it is expected that new degrees of freedom will emerge to play important role.
Otherwise inaccessible at the present energy scale, these fields might
interact with degrees of freedom of the standard model leading to some 
interesting theoretical predictions and observational signatures. Since string theory is a candidate
for a unified description of field interactions even upto the Planck scale,
we envisage that nature and the specific form of interaction of new fields with known degrees of freedom 
can be extracted from this theory in an unambiguous way. 
In this paper, we shall look for gauge invariant interactions of gauge fields
(electromagnetic, gravitational and $2$ and $3$-form gauge fields) to torsion.
In string theory, since the Kalb-Ramond (KR) field acts as a source
of torsion, we shall have a look at possible gauge and gravitational interactions 
of a this KR field. The KR  field is generic to any closed string spectrum but is {\it
not} a degree of freedom of the standard model. One can anticipate that any observational effect 
involving the KR field, obtained using standard fields as probes, is then a window into the
otherwise inaccessible world of very high energy physics supposedly
predicted by string theories. On the other hand, loop quantum gravity (LQG)
is also a candidate for quantum theory of gravity. 
In LQG, the Barbero-Immirzi parameter is a one-parameter
ambiguity which describes various topological sectors. This parameter also
comes up in the area spectrum and consequently in entropy of black holes wherefrom
its value is ascertained by comparing with the Bekenstein-Hawking entropy formula. If the Barbero-Immirzi
parameter is promoted to a field, it acts as a source for torsion. It is then interesting to
compare and contrast various interactions of fields with these two sources of torsion
that arise in these two theories of quantum gravity. Since there are observational implications,
the issue is even more satisfying.

In the context of the heterotic string theory, electromagnetic and 
gravitational interactions of KR fields arise quite naturally from
the requirements of consistency. As is well known \cite{gsw}, the ($E_8 \otimes E_8$)
or $SO(32)$ heterotic strings are two anomaly free gauge groups
which can be coupled to $N=1$ supergravity in $10$ dimensions. Anomaly
cancellation (the Green-Schwarz mechanism) requires that the KR 3-form
field strength is augmented by addition of ($E_8 \otimes E_8$)
Yang-Mills Chern-Simons 3-form and local Lorentz Chern-Simons
3-form \cite{gsw}. This augmentation induces electromagnetic and gravitational
interactions of the KR field which lead to potentially interesting
physical effects showing up in the Maxwell and Einstein equations,
when the theory is compactified to four dimensions. The
electromagnetic effect mainly comprise a rotation of the polarization
plane of electromagnetic waves from large redshift sources, upon
scattering from a homogeneous KR background \cite{ms,kmss,kmss2,djm,ih}.
This rotation is independent of the wavelength of the electromagnetic 
wave and cannot be explained by Faraday effect where the plane of
polarization of the electromagnetic wave rotates depending
quadratically on the wavelength while passing through some
magnetized plasma. Similarly, the gravitational interaction leads 
to the result that the plane of polarization of gravitational 
waves rotate through an angle that is proportional to (a
power of) the KR field strength component \cite{cm}. Predictions 
of this kind can then be useful if some deviations from the traditional expectations
are observed. For example, such interactions have been studied
within the framework of the five dimensional Randall-Sundrum braneworld model. When compactified to
four dimensions, they
lead to huge deviations from the expected results \cite{cm,mas,msenss,mss,msens} which can be used to put bounds on the 
various parameters in the theory \cite{mrs,agjkr,muksens}. On the other hand, if the predictions are non-observable,
they lead to upper bounds on the presence of new fields which is
important in our search for new theories and their couplings. To exemplify,
in the present case of rotation of plane of polarization of electromagnetic 
waves, the magnitude of the effect is sensitive to the dimensional
compactification of the underlying theory. For toroidal 
compactification (as well as for the Calabi-Yau compactification) of
the theory (in the zero slope limit), the predicted rotation is
proportional to the appropriate KR field strength component (scaled by
the inverse scale factor in a Friedmann universe), so that bounds on
the observed rotation translate into a stringent upper bound on the
size of the KR field strength component. Moreover, if one uses the bounds
on the KR field strength obtained  from the cosmic optical activity, the 
order of magnitude of the similar effect for gravitational waves can be calculated.

The interactions which give rise to the above-mentioned predictions
arise very naturally in string theory and they have been well studied.
Interestingly, one can also perceive of another class of interactions
which has not been discussed in this context except for in \cite{m,mms},
where only the electromagnetic interaction was considered. In this paper, we shall
extend the study to non-Abelian gauge fields and discuss
the effects of these possible new interactions in detail. Let us 
discuss the motivation for introducing such structures in brief 
(details will be in section $2$). The issue originally arose during the study
of Einstein-Cartan (EC) spacetime. The idea was to construct a 
gauge invariant coupling of electromagnetic field $(A_{\mu})$ to torsion- which 
is another geometrical property of the EC spacetime along with the metric.
The field strength ($F_{\mu\nu}$) for such a spacetime also depends
on torsion \cite{h}. However, because the torsion does not have a transformation under 
$U(1)$ gauge transformation, the electromagnetic field strength is not gauge 
invariant. This is dissatisfactory since we expect that field strengths
must be measurable even in spacetimes with torsion. This requirement on the field strength
demands that the torsion must also stay invariant under $U(1)$ gauge transformation.
This situation implies that there is a non-gravitational field,
possibly massless, to function as the source of the torsion \cite{ms}. Since that field
must be bosonic, one can opt for the KR antisymmetric second rank tensor field $B_{\mu\nu}$
as a possible candidate. $B_{\mu\nu}$, being a massless
antisymmetric field, is expected to be a gauge connection, as indeed it is, with the following gauge transformation
$\delta_{\lambda}B_{\mu\nu}=\partial_{[\mu}\lambda_{\nu]}$ and this leaves its field strength $H_{\mu\nu\lambda}$ gauge invariant.
Moreover, for anomaly-free quantum theory, $H_{\mu\nu\lambda}$ must be modified 
with the addition of an electromagnetic Chern Simons three tensor and
if $B_{\mu\nu}$ is endowed with a non-trivial electromagnetic gauge transformation along with 
Kalb-Ramond gauge transformations, the KR field strength remains invariant under $U(1)$ gauge transformation.
This is precisely what was needed: the torsion field is gauge invariant. Interactions of this type
gives rise to interactions in the form of rotation of plane of polarization of electromagnetic
(and gravitational) waves as discussed in the previous paragraph.

What if one wants to couple a $2$-form or a $3$-form gauge field to torsion? Such fields
arise in the perturbative and non-perturbative sector (D-branes) of string theory compactified
to four dimensions and in supergravity. Again, field strengths for such higher rank tansor field
are also not invariant under their respective gauge transformations in presence of
spacetime torsion. Once we take the KR field as a source for torsion,
there is a possible way out. We again demand the field strengths of $2-$ form or  $3$-form gauge
fields to be observable so that one again has to modify $H_{\mu\nu\lambda}$, but is a peculiar way. This extra term,
instead being of the form $A\wedge F$ for the ($U(1)$) case above, is $A\wedge {}^{*}F$, where ${*}$ denotes the Hodge
dual and $A$ is a one, two or a three form field. Again, if the field $B_{\mu\nu}$ has a non-trivial transformation
under the gauge transformation of the form fields, its field strength ($H_{\mu\nu\lambda}$) and hence torsion
remains invariant under gauge transformations, as required (for this case, we shall
work in order $O({\sqrt{G}})$).  It is also interesting to note that 
addition of such terms ($A\wedge {}^{*}F$) not only works for $2$ and $3$ form fields, but also for a $1$-form field.
Moreover, one gets an additional set of interaction for the electromagnetic fields and $H_{\mu\nu\lambda}$ field
with observable consequences. These issues were first discussed in \cite{m} and a possible embedding of such terms in $N=1$ supersysymmetric
theory was discussed in \cite{mms}.
In this paper, we shall extend the formalism for non-Abelian gauge fields and also for gravity waves and look for observational
predictions. Interestingly, because of the presence of the Hodge dual, interactions of the later kind violate
spatial parity. With the CMB data and the Planck data available, it might be interesting to look for such
ideas now. Indeed, observational implications of such terms have already been
discussed \cite{kamin, seto, kaminglu, wu, pos}. However, basis of terms 
have not been discussed in details and the coupling constant for such interactions are usually not pinned down.

The interest in LQG for such interactions and consequently it's relation or differences
with string theory/supergravity is due some recent studies \cite{tavyu,mer,tav,keto,cianmon, krasgo,ysoa}. These papers 
deal with the consequences of promoting the Barbero-Immirzi (BI) parameter to a field. It turns out that
the derivative of the BI field is the source for torsion. Moreover, since the BI field
is pseudo-scalar\footnote{The expression for area spectrum in LQG depends on the BI parameter and as such must be a pseudo-scalar
for a well-defined transformation property of the area element.}, it is natural to compare and contrast
this BI field with the axion \cite{mer}. If the BI field is an axion, it's derivative is dual to the $H_{\mu\nu\lambda}$ field 
alluded to  above and such fields might have interactions with electromagnetic and gravitational fields
in the way very similar to the one discussed above in the context of string theory. We shall discuss these issues in detail below
and point out to some observational implications.

The plan of the paper is as follows: In section 2, we discuss the gauge invariant
coupling of various form fields to torsion and show how this can be achieved
with special reference to electromagnetism and gravity. In the next section, we shall
review the consequences of such interaction for the Maxwell fields and extend them 
to gravity in the next section. In section 5, we compute the quantum effective-potential (Coleman-Weinberg potential) \cite{cw}
for a theory of gravity by including the modified interactions. We will see that inclusion of a parity violating scalar 
field (the BI field/axion) doesn't have any effect in the one-loop effective 
potential of a theory where higher curvature terms are present. We
conclude in section 6.

\section{Gauge Invariant Interactions of fields with torsion}

In the standard Einstein-Maxwell theory, 
the electromagnetic field-strength, reduces to the flat space expression on account of the symmetric nature of
the Christoffel connection. However, in the theory of gravity described by Einstein-Cartan theory,
\emph{i.e.} in case where one has spacetime torsion, the situation changes
quite drastically, because the electromagnetic field strength is no longer
gauge invariant \cite{h}. Indeed, it is easy to see that
\begin{equation}
F_{\mu\nu}=\partial_{[\mu}\,A_{\nu]}-T_{\mu\nu}{}^{\rho}\, A_{\rho},
\end{equation}
where, $T_{\mu\nu}{}^{\rho}\, A_{\rho}$ is the torsion (antisymmetric combination of the Christoffel connection), is obviously
not invariant under $U(1)$ gauge transformation $\delta_{\lambda}\,A=d\lambda$, $\lambda$ being the gauge function.
Since $F_{\mu\nu}$ and any field strengths must be measurable quantities even in a curved spacetime with torsion,
the torsion tensor, a purely geometric quantity like curvature must also be gauge invariant. However, this implies
that one must also have another geometrical quantity which might compensate for the loss of gauge invariance due
to torsion. In absence of such compensating fields, it is natural to look for non-gravitational fields to 
act as a source for torsion \cite{ms}. In the context of string theory, the Kalb-Ramond (KR) field seems to be an ideal
candidate source \cite{ms}. Indeed, it also has all the desired gauge transformation properties required of torsion.

In this section, we shall first review the basic facts about the KR field
as is known from string theory with special emphasis on it's gauge transformation
properties. The KR field is characterized by a 2-form potential $B$ which has a
3-form field strength $H \equiv dB$; the field strength is invariant
under the KR gauge transformation $\delta_{\bar{\lambda}} B~=~d \bar{\lambda}$,
where $\bar{\lambda}$ is a one-form gauge parameter. Immediately, one obtains
the Bianchi identity for the KR field:
\begin{eqnarray}
dH ~=~0 \label{krbi}
\end{eqnarray}
In  $4$ dimensional spacetime, the free KR action is
given by
\begin{eqnarray}
S_{H}~=~\int_{{\cal M}_4} ~H~\wedge ~^*H~, \label{kract}
\end{eqnarray}
where, $^*H$ is the Hodge-dual of the field strength $H$. Varying this
action w.r.t. $B$ yields the KR field equation
\begin{eqnarray}
d^*H~=~ 0 \label{kreom}
\end{eqnarray}
which has the local solution
\begin{eqnarray}
^*H~=~d \Phi_H, \label{krsol}
\end{eqnarray}
where, $\Phi_H$ is a scalar. Substituting this in the 
one obtains for the field $\Phi_H$
\begin{eqnarray}
d ^*d\Phi_H ~=~0~. \label{boxv}
\end{eqnarray}
Thus, on-shell the Bianchi identity for the field $B$ is the equation of motion
for it's Hodge dual field. This is not surprising and is a feature of
all Hodge-dual related fields.
 
Let us now point to the string theory connection.
$B$ occurs in the massless spectrum of the free string
in ten dimensional heterotic string theory. In the zero slope
limit, this theory reduces to ten dimensional $N=1$ supergravity coupled
to $N=1~ E_8 \otimes E_8$ super-Yang-Mills theory. The requirement of
ten dimensional supersymmetry and that the quantum theory be free
of all anomalies implies that the KR field strength $H$
be augmented as \cite{gsw}
\begin{eqnarray}\label{Haugmented}
H~=~dB~-~\frac{1}{M_{P}}~(\Omega_{YM}~-~\Omega_{L}) ~,\label{krfu}
\end{eqnarray}
where
\begin{eqnarray}
\Omega_{YM}~\equiv~\mbox{tr}(A \wedge dA~+~\frac23 g\, A\wedge A  \wedge A)
~\label{ymcs}
\end{eqnarray}
is the Yang-Mills Chern-Simons 3-form with $A$ the gauge  connection
1-form and $ M_{P}$ is the Planck mass in $4$- dimensional spacetime.
$\Omega_{L}$ is the gravitational Chern-Simons 3-form  obtained
by replacing the Yang-Mills gauge connection $A$ by the  spin
connection 1-form $\omega$, and the trace is  taken
over the local Lorentz indices. The augmentation in eq. (\ref{krfu}) has important consequences.
The field $H$, being a field strength, must remain gauge invariant under both Yang-Mills
gauge transformations and under local Lorentz transformations. This implies that $B$
must  now transform non-trivially under both gauge transformations inspite of 
$B$ being neutral. To simplify and to set the notations for the remaining part
of the paper, let us say that the gauge field $A$ is $U(1)$ valued. Then, the transformation of
$A$ is given by
\begin{eqnarray}
\delta_{\lambda} A ~=~ d\lambda \label{ymgt},
\end{eqnarray}
where, $\lambda$ is the gauge parameter. 
The Chern-Simons term now only contains $A\wedge dA$. We shall now
denote $\Omega_{YM}$ by $\Omega_{EM}$ and this term varies as 
\begin{eqnarray}
\delta_{\lambda}\, \Omega_{EM} ~=~ d \lambda~ \wedge  dA\label{ymcsvar}
\end{eqnarray}
Thus, to achieve gauge invariance for the $H$ field, the
transformation law for $B$ should include the $2$-form in (\ref{ymcsvar})
so that under Yang-Mills gauge transformation
\begin{eqnarray}
\delta_{\lambda}\, B = -\frac{1}{M_{P}}(\lambda~ dA) \label{ymb}
\end{eqnarray}

Also, the gravitational field in the vielbein formalism can
be treated very similarly to the Yang-Mills field. Specifically
the Yang-Mills potential $A$ is analogous to the spin connection
1-form $\omega_{AB}$, where $A, B$ are Lorentz indices. Under an
infinitesimal Lorentz transformation with parameters given by
an $SO(D- 1,1)$ matrix $\Theta$, the transformation of $\omega$
is
\begin{eqnarray}
\delta_{L} \omega  ~=~ d\Theta ~+~ [\,\omega,\Theta] \label{lgt},
\end{eqnarray}
The Lorentz Chern-Simons term varies as 
\begin{eqnarray}
\delta_{L} \Omega_{L} ~=~ \mbox{tr}(d \Theta~ \wedge d\omega)\label{lcsvar}
\end{eqnarray}
Similar to the argument above, transformation law for $B$
should include the $2$-form in (\ref{lcsvar})
so that under Lorentz transformation
\begin{eqnarray}
\delta_{L} B = - \frac{1}{M_P}\mbox{tr}(\Theta~ d\omega) \label{lmb}
\end{eqnarray}

Retaining the form of the KR action (\ref{kract}), it follows
that the KR field equation does not change. Therefore, $^*H$  still
has the local solution (\ref{krsol}). However, the KR Bianchi identity
certainly changes, leading to
\begin{eqnarray}\label{phieom}
d ^*d\Phi_{H}~=~\frac{1}{M_{P}}~\mbox{tr}(F \wedge F~-~R \wedge R) ~,
\end{eqnarray}
where $F(R)$ is the Yang-Mills (spacetime) curvature 2-form.
The Yang-Mills and Einstein equations change non-trivially. We
shall consider these below in special situations viz., the Maxwell
part of the gauge interaction and linearized gravity.

This scenario works well for $1$-form gauge fields. How about if we want a gauge invariant coupling of higher form fields to torsion?
In \cite{m}, it was proposed that one needs additional terms to be augmented to the KR field strength. For $U(1)$ gauge fields,
it was proposed that the following additional augmentation is needed (the argument is obviously not based on any requirements arising 
from string theory)
\begin{equation}\label{hmoreaug}
H\rightarrow H+\frac{1}{M_{P}}(A\wedge {}^{*}F)
\end{equation}
However, the origin of such terms remained obscure. In the appendix (see section \eqref{app}), we indicate the origin of such terms in this context. It is also
clear that in presence of such terms, the gauge transformation of $B$ field changes from that obtained in equation \eqref{ymb} \footnote{An immediate 
consequence of this gauge transformation is that the $H_{\mu\nu\lambda}$ now can no longer be thought of as a
parity eigenstate, and thus neither is its dual $\Phi_{H}$. In other words, one can decompose $\Phi_{H}=\Phi^{(+)}_{H}+\Phi^{(-)}_{H}$ where $+$ 
indicates even parity and $-$ is for odd parity. However, we shall continue to use the generic term $\Phi_{H}$ for this field.}:
\begin{equation}
\delta_{\lambda}\, B=-\frac{1}{M_{P}}(\lambda\, F+\lambda{}^{*}F)
\end{equation}
We can also proceed further and add to equation \eqref{hmoreaug} the spin-connection terms so that the augmentation  takes the following form:
\begin{equation}\label{hall}
H\rightarrow H+\frac{\zeta}{M_{P}}(\,A\wedge {}^{*}F+\,\omega \wedge {}^{*}R),
\end{equation}
where, $\zeta$ is a parameter which takes values $+1$ or $-1$. We have introduced this parameter since we donot quite fix the coefficient.
Now, instead of equation \eqref{phieom}, the result of such additional terms in equation \eqref{hall} is (we consider terms only upto
order $M_{P}^{-1}$)
\begin{eqnarray}\label{phineweom}
d ^*d\Phi_{H}~=~\frac{1}{M_{P}}~\mbox{tr}(F \wedge F~+\,\zeta\,F\wedge {}^{*}F-~R \wedge R -\,\zeta\,R\wedge{}^{*}R\,) ~,
\end{eqnarray}

In short, the upshot of the above analysis is that one can consider a
gauge invariant action of the following form \cite{ms,kmss}:
\begin{equation}\label{effectaction1}
S[g,T]=\int_{M_{4}} d^{4}x \,[\,R(g,T)-\frac{1}{4}F_{\mu\nu}F^{\mu\nu}-\frac{1}{2}H_{\mu\nu\lambda}H^{\mu\nu\lambda}+T_{\mu\nu\lambda}H^{\mu\nu\lambda}\,]
\end{equation}
where $H_{\mu\nu\lambda}$ is defined through equation \eqref{Haugmented} and the torsion tensor $T_{\mu\nu\lambda}$ is an auxilliary
field satisfying the constraint $T_{\mu\nu\lambda}=H_{\mu\nu\lambda}$. Putting the local solution $H=-{}^{*}d\Phi_{H}$ from equation
\eqref{krsol} in the action \eqref{effectaction1}, we get the effective equation for the field $\Phi_{H}$:
\begin{eqnarray}\label{effectaction}
S[g,A,\Phi_{H}]=\int_{M_{4}} d^{4}x\, [\,R(g,T)-\frac{1}{4}F_{\mu\nu}F^{\mu\nu}&-&\frac{1}{2}\partial_{\mu}\Phi_{H}\,\partial^{\mu}\Phi_{H}\,]\nonumber\\
                &+&\Phi_{H}(\,F \wedge F~+\,\zeta\,F\wedge {}^{*}F-~R \wedge R -\,\zeta\,R\wedge{}^{*}R\,)
\end{eqnarray}
which is precisely the action for a pseudo-scalar ($\Phi_{H}$) coupled to gravity\footnote{Now, because $\phi_{H}$ can be both parity violating
as well as parity conserving, each interaction is both parity conserving and parity violating. In what follows, we shall 
only consider the case where $\Phi_{H}$ is parity violating.}. Note that the extra interaction contributes
to the action in case of electromagnetism while is a higher derivative term for gravity. Without the $\Phi_{H}$ term, the higher
 derivative gravity terms $R\wedge \,R$ and $R\wedge\, {}^{*}R$ are the Pontryagin and the Euler invariants. They
are related to the gravitational axial current anomaly and stress-tensor anomaly respectively \cite{e,d,dc}. The equation of motion for this
pseudo-scalar is however given by equation \eqref{phineweom}. If the Barbero-Immirzi parameter is promoted to a field, the torsion is 
dual to the derivative of that pseudo-scalar field (just like the equation \eqref{krsol}). In that case, one gets an effective action
same as the first part of the action above \cite{krasgo,tavyu}. In the following sections, we study the consequences of such interactions.


\section{Electromagnetic interactions of KR field}
In this section, we shall confine our study to the electromagnetic
interactions of the KR field in four dimensional Minkowski
spacetime. Let us first restrict ourselves to the interaction of the type $\Phi_{H}F_{\mu\nu}{}^{*}F^{\mu\nu}$. Observe
that since the field $\Phi_{H}$ is a pseudo-scalar, the interaction is parity conserving. The relevant four dimensional field equations  are
\begin{eqnarray}
\partial_{\mu} H^{\mu \nu \rho}~& =&~0~ \nonumber \\ \partial_{\mu}
F^{\mu \nu}~&=&~ M_P^{-1}~H^{\nu \rho \eta}~F_{\rho  \eta}
~. \label{eom}
\end{eqnarray}
The corresponding Bianchi identities are
\begin{eqnarray}
\Box \Phi_H ~&=&~ M_P^{-1}~F^{\mu \nu}~^*F_{\mu \nu} \nonumber \\
\partial_{\mu} ^*F^{\mu \nu}~&=&~0 ~.\label{bia}
\end{eqnarray}
To simplify, let us assume that the 'axion' field $\Phi_H$ is {\it
homogeneous} and provides a background with which the Maxwell field
interacts. We restrict our attention to lowest order in the inverse
Planck mass $M_P$, so that terms on the RHS of the axion field
equation (\ref{bia}) are ignored to a first approximation.
Consequently, ${\dot \Phi}_H \equiv d\Phi_H/dt = f_0$ where  $f_0$ is
a constant of  dimensionality of $(mass)^2$. Under these conditions,
the Maxwell  equations can be combined to yield the inhomogeneous wave
equation  for the magnetic field ${\bf B}$
\begin{eqnarray}
\Box {\bf B}~=~-~{2~f_0\over M_{P}}~{\bf \nabla \times B} ~. \label{wave}
\end{eqnarray}
With the ans\"atz for a plane wave travelling in the  $z$-direction,
${\bf B}({\bf x},t)= {\bf B}_0(t)~\exp  ikz$, we obtain, for the
left and the right 
circular polarization states $B_{0 \pm}  \equiv B_{0 x} \pm i B_{0 y}$,
\begin{eqnarray}
{d^2 B_{0 \pm} \over dt^2}~+~(k^2 ~\mp~ {2 f_0 k \over M_P})~B_{0
\pm}=~0  ~.\label{poleq}
\end{eqnarray}
Similarly, we can obtain the wave equation for the left and 
right circularly polarization states for the electric field
which has exactly the same form as that of magnetic field.
We concentrate on the equation for magnetic field as the conclusions
will be same for that of electric field. 
Thus, the right and left circular polarization states have different
angular frequencies (dispersion)
\begin{eqnarray}
\omega_{\pm}^2 = k^2 ~\mp~ {2k f_0 \over M_P} ~\label{emdisp}
\end{eqnarray}
so that over a time interval $\Delta t$, the plane of polarization
undergoes a rotation (for large $k$)
\begin{eqnarray}
\Delta \Psi_{op} \equiv |\omega_+ - \omega_-|~\Delta t ~\simeq ~2  {f_0
\over M_P}~\Delta t ~.\label{rota}
\end{eqnarray} 
If we assume, just like in the case of Faraday rotation, the
existence of a coherent electromagnetic wave over a time $\Delta
t$ (in general, cosmologically, any vector  perturbation tends to thermalize with
time scales typically smaller  than $\Delta t$), in FRW spacetime, the value of observed angle of rotation 
also incorporates the scale factor \cite{kmss}
\begin{eqnarray}
\Delta \Psi_{op} \equiv |\omega_+ - \omega_-|~\Delta t ~\simeq ~2  {f_0
\over a^2(t) M_P}~\Delta t ~, \label{rotf}
\end{eqnarray} 
where, $a(t)$ is the scale factor and $\Delta t$ is now to be taken as
the look-back time. This means that $\Delta \Psi = \Delta \Psi(z)$, 
where $z$ is the red-shift, and increases with red-shift. This rotation
also differs from the better-understood Faraday rotation in that it is
{\it achromatic} in the limit of high frequencies. Observationally, even
for large redshift sources, the angle of  rotation is less than a
degree, which imposes the restriction on the  dimensionless quantity
$f_0/M_P^2 < 10^{-20}$, In regard to astrophysical
observations of optical activity, it appears that there is no definite
evidence that the rotation of the plane of polarization travelling
over cosmologically large distance is not entirely attributable
to Faraday rotation due to magnetic fields present in the galactic
plasma \cite{jr}. However, it is therefore not unlikely that the axion field will
endow observable effect in CMB.

In contrast, if we consider only the extra augmentation, \emph{i.e} the interaction $\Phi_{H}\,F_{\mu\nu}F^{\mu\nu}$, the resulting
wave equation for the ${\bf B}$ field lead to entirely different results. Observe that this interaction violates spatial parity.
The wave equation is simple to determine:
\begin{equation}
\frac{d^{2}{\bf B}}{dt^{2}}- 2\nabla {\bf B} + \frac{\zeta}{M_{P}} f_{0}\frac{d{\bf B}}{dt} = 0
\end{equation}
which eventually leads to the following equation for the left/right circularly polarised
light \cite{m}:
\begin{equation}\label{emmodulation}
\frac{d^{2}B_{+(-)}}{dt^{2}} + \frac{\bar{f}_{0}}{M_{P}}\frac{dB_{+(-)}}{dt} + k^{2}dB_{+(-)} = 0 ,
\end{equation}
where, $\bar{f}_{0}=\zeta\,f_{0}$.
The effect of parity violation is confined to the second term, which signifies either an enhancement or an attenuation, of the intensity of the observed 
electromagnetic wave, depending on the sign of $\bar{f}_{0}$ \cite{m}. We shall not go into the details of this calculation. Instead, we shall
show that a similar effect also exists for gravity waves which might lead to some observational effects.


\section{Behaviour of gravitational waves} 

First, let us discuss in some detail the gravitational analogue of the rotation of plane of polarisation (optical activity, equation \eqref{emdisp}) discussed
above \cite{cm}. This arises due to the parity conserving
term of the form $\Phi_{H} \mbox{tr}(R\wedge R) $ in equation \eqref{phieom}. First note that the augmentation of $H$ in (\ref{krfu})
implies that the $\mbox{tr}(\,R \wedge R\,)$ term contributes an additional term
to the Einstein equation over and above the energy-momentum  tensor of
the KR field. Formally,
\begin{eqnarray}
{\cal G}_{\mu \nu} ~=~{8\pi \over M_P^2}T_{\mu \nu} ~+~{16\pi  \over
M_P^3}~{1 \over \sqrt{-g}}~{\delta  \over \delta g^{\mu \nu}}~\int
d^4x'~\sqrt{-g}(x')~\Phi_H(x')~  R_{\rho \lambda \sigma
\eta}(x')~^*R^{\rho \lambda \sigma \eta}(x')~,
\label{ee1}
\end{eqnarray}
where,
\begin{eqnarray}
T_{\mu \nu}~=~H_{(\mu| \tau \rho}~H_{\nu)}{}^{\tau \rho}~-~\frac16
g_{\mu \nu} H^2 ~. \label{tmn}
\end{eqnarray}

Since our focus is on gravitational waves, it is adequate to  consider
the Einstein equation in a linearized approximation. To this effect,
we decompose the metric $g_{\mu \nu} = \eta_{\mu \nu} +  h_{\mu \nu}$
with the fluctuation $h_{\mu \nu}$ being considered  small so that one
need only retain terms of $O(h)$ in the Einstein  equation. We further
impose on the fluctuations $h_{\mu \nu}$ the Lorenz gauge
$h_{\mu \nu},{}^{\nu} = \frac{1} {2}  h,{}_{\mu} $ .

In this gauge, the
linearized Einstein equation from \eqref{ee1} becomes
\begin{eqnarray}
-\Box~ h_{\mu \nu}~=~{16\pi \over M_P^2}~T_{\mu \nu}~-~{128\pi \over
  M_P^3}~\epsilon_{(\mu|}^{~~~\sigma \alpha \beta}~\left[\Phi_{H,\lambda
  \sigma}~\left(h_{\beta |\nu), \alpha}^{~~~~~~\lambda}~+
  ~h^{\lambda}_{~\beta,\alpha |\nu)} \right) ~- ~\Phi_{H,\alpha}~\Box
  h_{\beta |\nu),\sigma} \right] ~ \label{linee}
\end{eqnarray}
Again we regard the axion field
$\Phi_H$ as a homogeneous background satisfying eq. (\ref{bia}) and
consider its effect on a plane gravitational wave and we restrict to  the
lowest inverse power of the Planck mass for which a nontrivial  effect
is obtained. We ignore terms on the RHS of the  axion field
equation
\begin{eqnarray}
\Box \Phi_H ~&=&~ M_P^{-1}~R^{\mu \nu \lambda \sigma}~^*R_{\mu \nu
\lambda \sigma }
\end{eqnarray}
We have chosen the Lorenz gauge and not all components
of  $h_{\mu \nu}$ are independent.
In fact, the only  physical degrees
of freedom of the spin 2 field are contained in  $h_{ij}$, for which
we choose a plane wave ans\"atz travelling in the z- direction,
\begin{eqnarray}
h_{ij}~=~\varepsilon_{ij}(t)~\exp ~-ikz ~. \label{gw}
\end{eqnarray}

The Latin indices above correspond to spatial directions. The other
components of $h_{\mu\nu}$ can be gauged away, so that  their field
equation need not be considered.
The only non-vanishing polarization components can be chosen to be
$\varepsilon_{11} = -~\varepsilon_{22}~,~\varepsilon_{12}=
\varepsilon_{21}$; from these the circular polarization components can
be constructed as in the Maxwell case: $\varepsilon_{\pm} \equiv
\varepsilon_{11} \pm i \varepsilon_{12}$. Further, 
we write the energy momentum tensor in eq. (\ref{tmn}) in terms of
$\Phi_{H}$ using eq. (\ref{krsol}). Then, under the approximation
of homogeneous axion field, these polarization
components satisfy the  inhomogeneous differential equation
\begin{eqnarray}
\left[{d^2 \over dt^2}~+~k^2 ~+~{\cal F}_{\pm} \right]
\varepsilon_{\pm}~=~-~{\cal F}_{\pm}~, \label{dife}
\end{eqnarray}
where,
\begin{eqnarray}
{\cal F}_{\pm}~\equiv~{8\pi f_0^2 \over M_P^2~(1 \pm  {128\pi k f_0
/M_P^3}) } ~ . \label{eff}
\end{eqnarray}
The difference between (\ref{dife}) and the analogous equation
(\ref{poleq}) is that the former has a forcing term absent in the
latter; this forcing term is dependent on the wave number $k$ and
controlled by the constant $f_0$ which characterizes the strength of
the KR field coupling. We are interested in large $k$, but we would still remain
within the Planckian regime $k < M_P$ so that the quantity $16\pi k
f_0/M_P^3 << 1$ and can serve as an expansion parameter, leading to
\begin{eqnarray}
\left[{d^2 \over dt^2} ~+~ k^2 ~+~ 8\pi f_0^2/M_P^2  ~\mp~1024\pi^2 k
f_0^3 /M_P^5 \right] \varepsilon_{\pm}~\simeq~-~8\pi f_0^2~(1 \mp 16
\pi k f_0/M_P^3)/M_P^2~ . \label{diffe}
\end{eqnarray}
We can now read off the dispersion relation
\begin{eqnarray}
\omega_{\pm}^2 ~ = ~ k^2 ~+~ 4\pi f_0^2/M_P^2  ~\mp~1024\pi^2 k f_0^3
/M_P^5 \label{disp}
\end{eqnarray}
whence the group velocity is $
v_{g \pm}~~=~ 1 ~+~ O(k^{-2})~$ and the phase velocity is given by $
v_{p \pm}~~=~ 1 ~\mp~512 \pi^{2}
(f_0^3/M_P^5k) ~$
As in the electromagnetic case, the rotation of the
polarization plane for gravitational waves is given by
\begin{eqnarray}
\Delta \Psi_{grav}~\simeq ~1024 \pi^{2} {f_0^3 \over M_P^5}~\Delta t ~.
\label{grota}
\end{eqnarray}
With the limits on $f_0$
given in the previous subsection, it is very small $O(10^{-30})$. However, since the tensor perturbations characterizing
the gravitational wave do not get randomized, so the effect is in
principle observable.


Let us now restrict ourselves to the parity violating term of the form $\Phi_{H}\mbox{tr}(R
\wedge {}^{*}R)$. The electromagnetic analogue of this term has been discussed in \cite{m,mms} and reviewed in
equation \eqref{emmodulation}. In contrast to the rotation of plane of polarisation for gravity waves
as observed above, equation \eqref{disp}, we expect some new consequences. In fact, we expect to observe modulation
for gravity waves. First, the effective action can be written as:
\begin{eqnarray}
{\cal G}_{\mu \nu} ~=~{8\pi \over M_P^2}T_{\mu \nu} ~+~{16\pi  \over
M_P^3}~{1 \over \sqrt{-g}}~{\delta  \over \delta g^{\mu \nu}}~\int
d^4x'~\sqrt{-g}(x')~\Phi_H(x')~  R_{\rho \lambda \sigma
\eta}(x')~R^{\rho \lambda \sigma \eta}(x')~,
\label{ee}
\end{eqnarray}
where,
\begin{eqnarray}
T_{\mu \nu}~=~H_{(\mu| \tau \rho}~H_{\nu)}{}^{\tau \rho}~-~\frac16
g_{\mu \nu} H^2 ~. \label{tmn1}
\end{eqnarray}
Again, we shall assume that the scalar field is homogeneous and it has only time dependence so that $d\Phi/dt=:f_{0}$ is a constant.
The spin $2$ field has only two degrees of freedom and the physical degrees of freedom are only
contained in $h_{ij}$. The equation of motion for the $h_{ij}$ can be determined in a straightforward manner:
\begin{equation}
\Box h_{ij}=-\frac{16\pi}{M_{P}^{2}}[-(\eta_{ij}+h_{ij})f_{0}^{2}\,]-\frac{16\pi}{M_{P}^{3}}\,\zeta\,[\,f_{0}\,\Box h_{ij,\,t}+\Phi_{H}\Box\,\Box h_{ij}\,]
\label{gwbapp}
\end{equation}
We choose a plane wave ans\"atz for $h_{ij}$, travelling in the $z-$ direction:
\begin{equation}
h_{ij}=\epsilon_{ij}(t)~exp{-ikz}
\end{equation}
As in the previous case, we shall assume that the only non-vanishing components of polarisation are $\epsilon_{11}=-\epsilon_{22}$ and 
$\epsilon_{12}=\epsilon_{21}$. Now, to facilitate the calculation, let us make some 
simplified assumption and notations. First, as seen from the previous section, let us define the dimensionless quantity
$\alpha:=(f_{0}/M^{2}_{P})<<1$. Secondly,
we shall remain in the Planckian regime but the wave number $k$ is such that the dimensionless quantity $\beta:=k/M_{P}$
is small (let us say $O(10^{-5})$). The modulas of the field $\Phi_{H}$ is taken to be order $1$. The previous equation now reduces to: 
\begin{equation}
\frac{d^{2}\epsilon_{ij}}{dt^{2}}+16\pi\,\alpha\zeta\,\beta\,k\frac{d\,\epsilon_{ij}}{dt}+
k^{2}(1-\frac{16\pi \alpha^{2}}{\beta})\,\epsilon_{ij}=\frac{16\pi\,f_{0}^{2}}{M_{P}^{2}}\,\eta_{ij} \label{eqgwamp}
\end{equation}
This is an equation for a damped oscillator with a forcing term. The system can get damped or can sustain gravity waves.
This depends on the value of the "$(b^{2}-4ac)$" term which here is given by:
\begin{equation}\label{discriminant}
2ik\,[1-\frac{16\pi\alpha^{2}}{\beta}+\frac{(16\pi\,\alpha\,\zeta\,\beta)^{2}}{4}]^{1/2}
\end{equation}
Let us list the various
possible cases. First, when $\alpha^{2}/\beta\,\ge 1$, \emph{i.e.} small values of $k$ (note that the third term in \eqref{discriminant}
is very small, with the value of $\beta$, it is of the order of $10^{-15}$ smaller compared to the second term and will not contribute appreciably), we get the scenario where the gravity waves dampen and is not observed:  
\begin{equation}
h_{ij}(t,z)=\mbox{exp}{(\,-\frac{16\pi\alpha\,\zeta k}{M_{P}}\,)}\,[A_{ij}\, e^{\bar{k}t-ikz}+B_{ij}\, e^{-\bar{k}t-ikz}]
\end{equation}

Second, consider the case when $\alpha^{2}/\beta\,< 1$ (\emph{i.e.} large values of $k$). Then, the solutions of the equation \eqref{eqgwamp} are:
\begin{equation}\label{eqnampatten}
h_{ij}(t,z)=\mbox{exp}{(\,-\frac{16\pi\,\zeta\,\alpha\, k}{M_{P}}\,)}\,[\,A_{ij}\, e^{ikt-ikz}+B_{ij}\, e^{-ikt-ikz}]
\end{equation}
This is the standard solution where the wave proceeds sinusoidally. It is clear that the solution to this equation can give attenuation/amplification 
of amplitude of gravity waves. To see this, choose $\zeta=+1$ then the equation \eqref{eqnampatten} leads to attenuation of gravity waves
whereas for $\zeta=-1$, we get amplification of gravity waves.  
In short, in this case we do not see any rotation of plane of polarisation of gravity wave, rather the 
attenuation/amplification of the wave during propagation is the result of such an interaction. Such phenomena for
gravity waves was suggested in \cite{kamin} which however was largely phenomenological. If such effects are present, 
they have implications for CMB spectrum. They lead to non-zero cross-correlation in multipole moments $C_{l}^{TB} ~\mbox{and}~C_{l}^{EB}$.
Such effects cannot be induced by Faraday
rotation (if there is any intervening magnetic fields). This is 
because it is an anisotropic effect which will also change $l$. With the Planck data coming up, we expect to see some
of these effects or if these are not seen, the experiments can be used to put bounds on the coupling constants for these interactions.

\section{Effective Potential for Higher derivative gravity}
In this section we will study the effects of quantum fluctuations of different fields for a theory governed by the action \eqref{effectaction} by calculating the one-loop effective potential
 using loop-expansion scheme \cite{jackiw}. We will concentrate on the gravitational part of the action only. Effective-potential serves
 as a useful tool to investigate the vacuum structure of such a theory where one can define the theory to be valid upto an energy
scale (Planck energy) through cut-off and make predictions treating it as an effective theory. To keep the matters very general, we shall 
consider a theory of gravitation coupled with three different kinds of matter fields. The Einstein term is minimally coupled with a 
massive/massless scalar field $\phi_S$ which has a self interacting potential. The action also contains an 
(axion) field $\phi_A$ coupled with a CP-odd term $R_{\mu\nu\alpha\beta}\,{}^*R^{\mu\nu\alpha\beta}$ and another field $\phi$ which is 
coupled to the CP-even term $R_{\mu\nu\alpha\beta}\,R^{\mu\nu\alpha\beta}$. In Euclidean signature, the Lagrangian of the theory is
\begin{eqnarray}
{\cal L}&=&\nonumber{\cal L}_{g1}+{\cal L}_{g2}+{\cal L}_{g3}+{\cal L}_m \nonumber \\
&=&-\frac{1}{\kappa^2}\, R + a\,\phi\,R_{\mu\nu\alpha\beta}\, R^{\mu\nu\alpha\beta}+ b\,\phi_A\,R_{\mu\nu\alpha\beta}\,{}^*R^{\mu\nu\alpha\beta}\nonumber\\ 
&+&\frac{1}{2}\,g^{\mu\nu}\,\partial_{\mu}\phi\,\partial_{\nu}\phi+\frac{1}{2}\,g^{\mu\nu}\,\partial_{\mu}\phi_A\,\partial_{\nu}\phi_A +
\frac{1}{2}\,g^{\mu\nu}\,\partial_{\mu}\phi_S\,\partial_{\nu}\phi_S + V(\phi_S) 
\end{eqnarray}
where $\kappa^2=16\pi G$ and  $a,b$ are coupling constants which can be specified later (they are $M_{P}^{-3}$). Let us now turn to 
calculate the effective potential. For that purpose,
we first expand the metric $g_{\mu\nu}$ around a flat background:
\begin{equation}\label{expan_ep}
g_{\mu\nu}=\delta_{\mu\nu} + \kappa\, h_{\mu\nu},
\end{equation}
where $\delta_{\mu\nu}$ is a flat background and the fluctuations $h_{\mu\nu}$ are small,  $|h_{\mu\nu}|<1$.
For the decomposition \eqref{expan_ep}, the inverse of the metric is
\begin{equation}
g^{\mu\nu}= \delta^{\mu\nu} -\kappa\,h^{\mu\nu} + \kappa^{2}\, h^\mu{}_\lambda \, h^{\lambda\nu}+ \dots
\end{equation}
Furthermore, the determinant of the metric, which will be needed in the following, will be given by:
\begin{equation}
(g)^{\frac{1}{2}}=1 + \frac{1}{2} h^\alpha{}_\alpha - \frac{1}{4}h^\alpha{}_\beta \, h^\beta{}_\alpha + \frac{1}{8}\bigl( h^\alpha{}_\alpha\bigr)^2 + 
\dots 
\end{equation}
To calculate one-loop effective potential we need to expand the Lagrangians only upto quadratic order in the $h_{\mu \nu}$. The expansions
are listed below:
\begin{eqnarray}
\sqrt{g}\,{\cal L}_{g1}= \sqrt{g}\, R  & = &-{1 \over 4} \partial_{\alpha} h_{\mu \nu} \,\partial^{\alpha}
h^{\mu \nu} + {1 \over 4} \partial_{\alpha} h \,\partial^{\alpha} h - {1 \over 2}\partial_{\alpha} h\,
\partial_{\beta} h^{\alpha \beta} \nonumber\\
& & +{1 \over 2}\partial_{\alpha} h_{\mu \beta}\, \partial^{\beta} h^{\mu \alpha}+ \mathrm{total\,\, derivatives} \label{L1expnsn}
\end{eqnarray}
The expressions for the other two terms are long. However, we give them below. First,
\begin{eqnarray}
\sqrt{g}\,{\cal L}_{g2}&=&\sqrt{g}\,a\,\phi\,R_{\mu\nu\alpha\beta}\,R^{\mu\nu\alpha\beta}\nonumber\\
&=& a\,\kappa^{2}\,(\partial_{\nu}\partial_{\rho}\phi\,h_{\mu\sigma}\,\partial^{\nu}\partial^{\rho}h^{\mu\sigma}+ 
\partial_{\rho}\phi\,h_{\mu\sigma}\Box\,\partial^{\rho}h^{\mu\sigma}+\phi\,h_{\mu\sigma}\Box\Box\,h^{\mu\sigma}\nonumber\\
&+& \partial_{\nu}\partial_{\rho}\phi\,h_{\mu\sigma}\partial^{\mu}\partial^{\sigma}h^{\nu\rho}
+\partial_{\rho}\phi\,h_{\mu\sigma}\,\partial^{\mu}\partial^{\sigma}\partial_{\nu}h^{\nu\rho}+
\phi\,h_{\mu\sigma}\,\partial^{\mu}\partial^{\sigma}\partial_{\nu}\partial_{\rho}h^{\nu\rho}-
2\,\partial_{\nu}\partial_{\rho}\phi\,h_{\mu\sigma}\,\partial^{\nu}\partial^{\sigma}h^{\mu\rho}\nonumber\\&-&
2\,\partial_{\rho}\phi\,h_{\mu\sigma}\Box\,\partial^{\sigma}h^{\mu\rho}-2\,\phi\,h_{\mu\sigma}\Box\,\partial^{\sigma}\partial_{\rho}h^{\mu\rho})\label{L2expnsn}
\end{eqnarray}
and
\begin{eqnarray}
\sqrt{g}{\cal L}_{g3}&=&\sqrt{g}\,b\,\phi_A\,R_{\mu\nu\alpha\beta}\,{}^*R^{\mu\nu\alpha\beta}\nonumber\\
&=&2b\,\kappa^{2}\,\left\{\partial_{\lambda}\partial_{\sigma}\phi_A\,\partial_{\alpha}\partial^{\lambda}\,h^{\rho}_{\beta}h_{\rho\eta}+
\partial_{\sigma}\phi_A\,h_{\rho\eta}\Box\,\partial_{\alpha}h^{\rho}_{\beta}-\partial_{\lambda}\partial_{\sigma}\,h_{\rho\eta}\partial_{\alpha}
\partial^{\rho}h^{\lambda}_{\beta}\right\}\epsilon^{\alpha\beta\sigma\eta}\label{L3expnsn}
\end{eqnarray}
Note that due to the presence of a Levi-civita tensor which is completely anti-symmetric in it's indices, only three terms will survive in 
the expansion of $\La_{g3}$,
Since we are calculating one-loop effective potential, terms of order $2$ in fluctuations will only contribute. To obtain one-loop effect, 
it is sufficient to choose spacetime independent saddle points for the scalar (and pseudo-scalar) fields;
\[\phi(x)=\phi_0+\Phi(x);\,\, \phi_A(x)=\phi_{A0}+\Phi_{A}(x);\,\, \phi_S(x)=\phi_{S0}+\Phi_S(x)\]

With these choices, the derivative terms of the scalar fields will not contribute to the resulting Lagrangian (expanded about the saddle points). 
The Lagrangian relevant for calculating one loop effective potential is by invoking the transverse-traceless gauge \cite{smolin,ovrut}. With 
$\partial_{\mu}h^{\mu\nu}=0$ and $h=0$, the relevant part of the lagrangian becomes: 
\begin{eqnarray}
\La_{rel}&=&\nonumber\frac{1}{4}h_{\mu\nu}(-\Box_{E})\,h^{\mu\nu}+a\,\kappa^{2}\,\phi_0\,h_{\mu\nu}\,\Box_{E}\,\Box_{E}\,h^{\mu\nu}-\frac{1}{2}
\Phi_S(-\Box_{E}+V''(\phi_{S0}))\Phi_S-V(\phi_{S0})\nonumber\\&-&\frac{1}{4}\,\kappa^{2}\,h_{\mu\nu}V h^{\mu\nu}+\frac{1}{2}\Phi\,(-\Box_{E})\Phi+
\frac{1}{2}\Phi_{A}\,(-\Box_{E})\Phi_{A} , \label{Lrelvnt}
\end{eqnarray}
where $\Box_{E}$ is the operator in Euclidean space.
Since we are perturbing around a flat background, ghost doesn't appear in this gauge \cite{cai}. Also, it is important to 
note here that the (axion) field $\Phi_{A}$ has no contribution to the one-loop effective potential. 
Now, eqn (\ref{Lrelvnt}) may be conveniently written as
\begin{equation}
{\cal L}_{rel}={1 \over 2}h_{\mu \nu}{\cal O}^{\mu\nu\alpha\beta}h_{\alpha\beta}+\frac{1}{2}\Phi_S(-\Box_{E}+V''(\phi_{S0}))\Phi_S+
\frac{1}{2}\Phi\,(-\Box_{E})\Phi+\frac{1}{2}\Phi_{A}\,(-\Box_{E})\Phi_{A}
\end{equation}
where the operator \[{\cal O}^{\mu\nu\alpha\beta}=\frac{1}{2}\delta^{\mu\alpha}\delta^{\nu\beta}\left[-\Box_{E}+ 2a\kappa^2\,\phi_0\,\Box_{E}\Box_{E}-
\kappa^2\,V(\phi_{S0})\right]\] 
Now, we rewrite the Lagrangian in terms $\Psi_{i}$ where $i=1,2,...10$ denotes ten independent components of $h_{\mu\nu}$ \cite{veltman}.
\begin{equation}\La_{rel}=\frac{1}{2}\Phi(-\Box_{E}+V^{''}(\phi_{S0}))\Phi
+\frac{1}{2}\Psi_iM_{ij}\Psi_j ,\label{eqnij}
\end{equation}
where we have we have employed the following index correspondence: $\mu\nu\rightarrow i$ and $\alpha\beta\rightarrow j$. 
To get the one-loop effective potential we need to calculate the determinants of differential operators which in this case reduces 
to calculate the eigenvalues of the $10 \times 10$ matrix $M_{ij}$ \cite{veltman}. The
operator for scalar field is trivial. We write down the eigenvalues,
\begin{eqnarray}
\lambda_i&=&-\frac{1}{2}(k^2+4a\kappa^2\phi_0k^4-\kappa^2V) \,\,;(1 \leq i \leq 4) \nonumber \\ 
\lambda_i&=&\left(k^2+4a\kappa^2\phi_0k^4-\kappa^2V\right)\,\,;(5 \leq i \leq 10) \label{egnvlues}
\end{eqnarray}
The one-loop effective potential is given by
 \begin{equation}
V_{\mbox{eff}}^{(1)}=V(\phi_{S0}) + {1 \over 2}\mbox{Tr}\ln(k^2+V^{''})+\sum_{i=1}^{10}\frac{1}{2}\mbox{Tr}\ln\lambda_i ~,
\end{equation}
where $\mbox{Tr}$ is the functional trace. Performing the momentum space integrals and introducing a cut-off we obtain the unrenormalized 
one-loop effective potential
\begin{eqnarray}
V_{\mbox{eff}}\,(\phi_{S0},\phi_0)&=&\frac{5}{16\pi^2}\left[\left(\frac{\Lambda^4}{2}-\frac{1-2eg}{4e^2}\right)\ln{\frac{e\Lambda^4}{g}}
+\frac{\Lambda^2}{2e}+\frac{g}{2e}-\frac{1}{4e^2}\right.+\left.\frac{\sqrt{1-4eg}}{4e^2}\ln{\left(\frac{1+\sqrt{1-4eg}}{1-\sqrt{1-4eg}}\right)}\right]
\nonumber\\&+&\frac{\Lambda^2\,V^{''}}{32\pi^2}
+\frac{V^{''2}}{64\pi^2}\left(\ln{\frac{V^{''}}{\Lambda^2}}-\frac{1}{2}\right)+V(\phi_{S0}) \label{UnRnmEPv1}
\label{effpv2}
\end{eqnarray}
where $e=4\phi_0a\kappa^2 \,\mbox{and}\,\, g=-\kappa^2V$, $\Lambda^2$ is the momentum cutoff. If we put the expressions 
of $e$ and $g$ back into the above expression the effective potential is seen to have an imaginary part: 
\begin{eqnarray}
V_{\mbox{eff}}\,(\phi_{S0},\phi_0)&=&\frac{5}{16\pi^2}\left[\left(\frac{1+8\kappa^4\phi_0aV}{64\kappa^4\phi_0^2a^2}
-\frac{\Lambda^4}{2}\right)\ln{\frac{V}{\Lambda^4}}+\frac{\Lambda^2}{8\kappa^4\phi_0a^2}-\frac{V}{2}-
\frac{1}{64\kappa^4\phi_0^2a^2}\right.
\nonumber\\&+&\left.\frac{\sqrt{1+8\kappa^4\phi_0aV}}{64\kappa^4\phi_0^2a^2}\ln{\left(\frac{1+\sqrt{1+8\kappa^4\phi_0aV}}{1-\sqrt{1+8\kappa^4\phi_0aV}}\right)}\right]+
\frac{5i}{16\pi}\left(\frac{1+8\kappa^4\phi_0aV}{64\kappa^4\phi_0^2a^2}-\frac{\Lambda^4}{2}\right)\nonumber\\&+&
\frac{\Lambda^2\,V^{''}}{32\pi^2}+\frac{V^{''2}}{64\pi^2}\left(\ln{\frac{V^{''}}{\Lambda^2}}-\frac{1}{2}\right)+V(\phi_{S0}) \label{UnRnmEPv11}
\label{effpv3}
\end{eqnarray}
It is interesting to see here that an imaginary part is generated in the effective potential. Similar kind of result was found 
in \cite{smolin} for a theory where a single scalar field is coupled to gravity. The imaginary part of the effective potential
signifies that we have chosen an unstable vacuum, in fact flat space is not a stable vacuum of this theory. The value of $V_{eff}$ at t
he asymmetric minimum serves as a cosmological constant at the tree level \cite{smolin, cho1}. This interpretation 
can be explained as follows: Let  $V_{eff}$ develops an symmetry breaking minima at the value 
of $\phi_{S0}=\phi_{S_{min}}$ and $V_{eff}(\phi_{S_{min}})\neq 0$ then $V_{eff}(\phi_{S_{min}})$ will act as a cosmological 
constant at the tree level. Now we include a cosmological constant to this theory, so now we have a different vacuum state not a 
flat space but a de- Sitter space. The Lagrangian reads as 
\begin{eqnarray}
{\cal L}&=&-\frac{1}{\kappa^2}\, (R -2C)+ a\,\phi\,R_{\mu\nu\alpha\beta}\, R^{\mu\nu\alpha\beta}+ b\,\phi_A\,R_{\mu\nu\alpha\beta}\,{}^*R^{\mu\nu\alpha\beta}\nonumber\\ 
&+&\frac{1}{2}\,g^{\mu\nu}\,\partial_{\mu}\phi\,\partial_{\nu}\phi+\frac{1}{2}\,g^{\mu\nu}\,\partial_{\mu}\phi_A\,\partial_{\nu}\phi_A +
\frac{1}{2}\,g^{\mu\nu}\,\partial_{\mu}\phi_S\,\partial_{\nu}\phi_S + V(\phi_S) \label{lagrwthC}
\end{eqnarray}
where $C$ is the cosmological constant. If we repeat the calculation for the effective potential from \eqref{lagrwthC}, 
the imaginary part of the potential will be
\begin{equation}
\mbox{Im}\,[V_{eff}(\phi_{S0},\phi_0)]=\frac{5}{16\pi}\left(\frac{1+2(\kappa^2V+2C)\,\phi_0a\kappa^2}{64\kappa^4\phi_0^2a^2}-\frac{\Lambda^4}{2}\right)
\end{equation}
It is now obvious that we can fine tune the cosmological constant $C$ such that the imaginary part of $V_{eff}$ and the cosmological constant
 both vanishes
\begin{equation}
{1\over2}\kappa^2V(\phi_{S_{min}})+\frac{1}{4\phi_0a\kappa^2}+C=0
\end{equation}
This makes the flat background a solution of the Einstein equation at the vacuum state.  
The calculation of effective potential here done in conventional approach which is not devoid of gauge ambiguities. 
However, it is well known that Vilkovisky-DeWitt (VD) \cite{vilkovisky,dewtt} approach of deriving effective potential is free 
from any ambiguities related to gauge-fixing condition or parameterization of the theory. We don't employ the method of VD here,
 although quite a number of papers have already been in the literature which calculate the effective potential in VD approach 
for ordinary and higher derivative gravity \cite{cho,odin,shapiro}. VD effective potential for the theory under consideration may be taken as a future project. 


\section{conclusions}

Let us first recall the results of the paper. In string theory, the Kalb-Ramond field acts as a source term for torsion which has 
various interactions with gauge fields. In order that the interactions are gauge invariant, the Kalb-Ramond field
$B_{\mu\nu}$ must be endowed with non-trivial transformations under gauge fields. This leads to some interesting interactions
with observable consequences. One of them is the rotation of plane of polarisation for electromagnetic and gravity waves.
These had been studied earlier and have been matched with experimental results. However, these interactions are not the only possible
ones. One can have additional ones which arise from the gauge invariant coupling of higher form fields to torsion. Such interaction
was proposed in \cite{m}. We give a theoretical basis for such terms and extend the formalism for gravity waves. Observational consequences 
of such interactions are alltogether different. They lead to amplification/attenuation of electromagnetic or gravity waves
and have important implications for anisotropy of the Cosmic Microwave
Background (CMB) by spatial parity violation \cite{kamin}.
For such parity breaking term, one can get certain non-vanishing multipole moment correlations between the temperature anisotropy and polarization
of the CMB. In the CMB data, one usually observes correlations like $C_{l}^{TT},C_{l}^{EE},C_{l}^{BB}\, \mbox{and}\,C_{l}^{TE}$ which 
arise from parity conserving interactions. On the other hand, cross-correlations like $C_{l}^{EB}\, \mbox{and}\,C_{l}^{TB}$ arise from
parity violating interactions from which bounds on the strength of such parity violating terms can be ascertained. We also study the Coleman-Weinberg
mechanism for such extended theory. This leads to a potential which might have some significance in the early universe and inflation. Initial
studies with this potential show that one can generate the requisite number of e-foldings from such a theory near the Planck scale. other consequences 
from such a potential requires further study.

\section{Appendix}\label{app}

In this appendix, we  shall show the existence of the extra term of the form $(A\wedge {}^{*}F)$ added to the 
KR field in equation . The question is: where to look for such terms?  To motivate, let us recall that the usual Chern-Simons term ($\Omega_{YM}$) 
augmented to the KR field strength $H$ in equation  is actually a boundary term. In the $U(1)$ version for example, the Chern-Simons term ($\Omega_{YM}$)
reduces to $(A\wedge F)$ which is precisely the conribution to boundary term corresponding to $(F\wedge F)$  
in $U(1)$ gauge theory. In the same token, let us look for the boundary terms for the action itself.
Consider the Lagrangian $4$-form for the \emph{free} Yang-Mills theory
\begin{equation}
L=\mbox{tr} (F\wedge *F)
\end{equation}
The on-shell variation of the Lagrangian gives 
\begin{equation}
\delta L= 2\mbox{tr}\,d(\delta\,A \wedge {}^{*}F):=d\Theta(\delta)
\end{equation}
The term $\Theta(\delta)$ is a three form and is often called the symplectic potential. Now, consider
the variation of the one-form $A$ through a parameter $\mu$, $0\le \mu\le 1$ and define:
\begin{eqnarray}
\delta_{\mu}\, A:= A\, \delta\mu  ~~~~ \mbox{and}~~~~ A_{(\mu)}:=\mu\,A  ~~~~\mbox{so that}\\
{}^{*}F_{(\mu)}=\mu{}^{*} F + (\mu^{2}-\mu){}^{*}(A\wedge A)
\end{eqnarray}
This implies that 
\begin{equation}\label{theta_var}
\Theta(\delta_{\mu})=2tr\,(A\wedge {}^{*}F_{(\mu)})\,\delta\mu
\end{equation}
Thus, on-shell, the above equation \eqref{theta_var} is equivalant to:
\begin{equation}
\frac{\delta}{\delta\mu}\mbox{tr} (F\wedge {}^{*}F)=2\,d\, tr\,[\mu \, A \wedge{}^{*} F + (\mu^{2}-\mu)A\wedge{}^{*}(A\wedge A) ]
\end{equation}
Integrating with respect to $\mu$, we get
\begin{eqnarray}
\mbox{tr} (F\wedge {}^{*}F)&=&d\, tr\,[A \wedge{}^{*} F -\frac{1}{3}\,A\wedge{}^{*}(A\wedge A) ]\\ \nonumber
&=&d\, tr\,[A \wedge{}^{*} dA+\frac{2}{3}\,A\wedge{}^{*}(A\wedge A) ]
\end{eqnarray}
Note that this term arises from a boundary contribution and is valid only on-shell. In contrast, the usual Chern-Simons term,
which can be derived in a similar fashion from the other boundary term $\mbox{tr}(F\wedge F)$ 
only requires the Bianchi identity. In standard treatments, the boundary term vanishes by the boundary conditions on the
fields. The above derivation is merely to show the existence of such terms in general when the field has all possible configurations.

Two comments are in order. Firstly, in the equation above, we have considered only the \emph{free} Yang-Mills theory. Now suppose that 
the Yang-Mills field is also coupled to other fields as the KR field $H_{\mu\nu\lambda}$ in the present paper. In that case, the equation of
motion for the Yang-Mills field is not merely
$D{}^{*}F^{i}=0$, but has contributions from the KR fields too. One then needs to look for the modification due to presence of such terms also. 
Secondly, as mentioned in the paper, we want not only to couple $1$-form field to $H$ field but also $2$ and $3$-form fields. In those 
cases, the term $[A\wedge {}^{*}(A\wedge A)]$ does not arise (and in not a $3$- form). For this reason, in what follows, we discard that term altogether.

Let us now comment on the effect of the equation of motion. From above construction, we get the following:
\begin{equation}
\mbox{tr} (F\wedge {}^{*}F)=d\, tr\,[A \wedge{}^{*} F ]+ \mbox{tr}(\delta A\wedge D{}^{*}F)
\end{equation}
For $\delta A^{i}=d\lambda^{i} +[A,\lambda]^{i}$, another term needs to be added to the first term. Thus in total, we get the
contribution to the total derivative to be:
\begin{equation}
\mbox{tr} (F\wedge {}^{*}F)=d\, tr\,[A \wedge{}^{*} F +\lambda D{}^{*}F ]
\end{equation}
To understand the effect of this term, let us restrict to $U(1)$ gauge theory for simplicity.
For $U(1)$ gauge fields, the effect of this augmentation leads to:
\begin{equation}
H\rightarrow H=dB +\frac{1}{M_{P}}(A \wedge{}^{*} F +\lambda d{}^{*}F )
\end{equation}
We want that $H$ remain gauge invariant under $U(1)$ gauge transformation. Then, $B$ must transform under $U(1)$ gauge transformation. This
can be easily found from the above equation:
\begin{equation}
\delta_{\lambda}\,B=\lambda\, {}^{*}F
\end{equation}
Thus, the gauge transformation of $B$-field comes out cleanly only when the contribution from the equation of motion is taken into account.
Then, why have we not added the term $\lambda d{}^{*}F$ in equation ? That is because we want to look only for effects of order $M_{P}^{-1}$
while the contribution of second term is of order $M_{P}^{-2}$.

\vspace{0.5cm}\noindent {\bf Acknowledgment}: We would like to thank Arindam Mazumdar for useful discussions on cosmological 
implications of this work. We also thank Sandipan Sengupta for some discussions.


\end{document}